\documentclass{article}
\usepackage[final]{colm2026_conference}
\usepackage[T1]{fontenc}
\usepackage{amsmath,amsfonts,bm}

\def\eqref#1{equation~\ref{#1}}
\def\1{\bm{1}}

\DeclareMathAlphabet{\mathsfit}{\encodingdefault}{\sfdefault}{m}{sl}
\SetMathAlphabet{\mathsfit}{bold}{\encodingdefault}{\sfdefault}{bx}{n}

\usepackage{graphicx}
\usepackage{booktabs}
\usepackage{array}
\usepackage{url}
\usepackage{microtype}
\usepackage{hyperref}
\definecolor{darkblue}{rgb}{0, 0, 0.5}
\hypersetup{colorlinks=true, citecolor=darkblue, linkcolor=darkblue, urlcolor=darkblue}
\usepackage{listings}
\usepackage{framed}
\usepackage{enumitem}

\title{Optimizing Expert-Designed Crystal Graph Networks for Band-Gap Prediction\\ with an Autonomous LLM Research Loop}

\author{
Chenmu Zhang \& Boris I. Yakobson \\
Department of Materials Science and NanoEngineering \\
Rice University \\
Houston, TX 77005, USA \\
\texttt{zcmben2014@gmail.com}
}

\begin{document}
\maketitle
% The venue is the LM4Sci workshop at COLM, not the COLM main conference; the
% [final] header is set inside \maketitle, so override it here.
\lhead{Accepted at the LM4Sci Workshop, COLM 2026}

\begin{abstract}
% NOTE: importance, maturity, opportunity, then the result (the most accurate no-external-pretraining model on the MatBench band-gap benchmark), then the analysis (only known methods, of two kinds, with two contributing examples), then the meaning and its limitation (no real methodological novelty).
Predicting a material's properties from its structure is a central, fast-advancing problem in
computational materials science. A decade of work has produced standard public benchmarks and many
published machine-learning models for the task \citep{dunn2020matbench}. The task's fixed metric and
these baselines make it a natural setting for autonomous agent research \citep{karpathy2026autoresearch}. On the MatBench band-gap
benchmark ($>$100k crystals), a general-purpose coding agent autonomously built the most accurate
model trained without external pretraining, ahead of all seventeen expert-designed models reported
for the task. A closer analysis shows it reached this by
implementing known methods: either already standard in crystal neural-network models, or borrowed
from other areas of machine learning. The contributing implementations include element-pair features
on each message-passing edge and a crystal space-group embedding. The work not only demonstrates that LLM-agent autonomous research can optimize an expert-designed
machine learning model for material property prediction, but also investigates the limitations of
such autonomous research.
\end{abstract}

\section{Introduction}
% NOTE: background, computational materials science only -- the problem, the standardized benchmarks and the body of published models, then graph networks as the most successful framework and three expert-designed techniques each motivated by a physical reason.
Predicting a crystal's properties from its atomic structure is a central problem in computational
materials science. A decade of work has produced standardized public benchmarks, from single-property
suites \citep{dunn2020matbench} to stability prediction built on machine-learned interatomic potentials
(learned models of a structure's energy) \citep{riebesell2023matbenchdiscovery}, together with a large
body of published models evaluated against them \citep{chen2019megnet,ruff2023cogn}. The most successful of
these frameworks are graph neural networks, which represent a crystal as atoms joined by bonds and so build
in the local geometry. These networks are expert-designed
from physical reasoning, for example: \citet{xie2018cgcnn} encoded each atom's local chemical environment in a crystal graph, \citet{schutt2018schnet} used continuous-filter
convolutions so a predicted energy varies smoothly as the atoms move, and \citet{choudhary2021alignn}
added a line graph to encode the bond angles that many properties depend on.

% NOTE: bridge from the mature benchmark to autonomous research -- these conditions suit an agent; recent autonomous research has succeeded but mostly in ML and CS, so whether an LLM coding agent can do materials autoresearch is open, because success there needs both correct code and scientific judgment.
A fixed metric and a field of expert-designed baselines are the conditions an autonomous research
agent is built to exploit. Autonomous, agent-driven research has advanced quickly, with recent systems running full machine-learning research pipelines end to end
\citep{lu2024aiscientist,yamada2025aiscientistv2}. That progress, however, has stayed
mostly within machine learning and computer science. Whether an LLM-driven coding agent can do the same on
a materials-science question has not been established. Success there needs two things: writing correct
code, which these agents handle increasingly well \citep{jimenez2024swebench,yang2024sweagent,li2022alphacode}, and the scientific judgment to turn
each result into the next modeling decision.

\begin{figure}[t]
\centering
\includegraphics[width=0.82\linewidth]{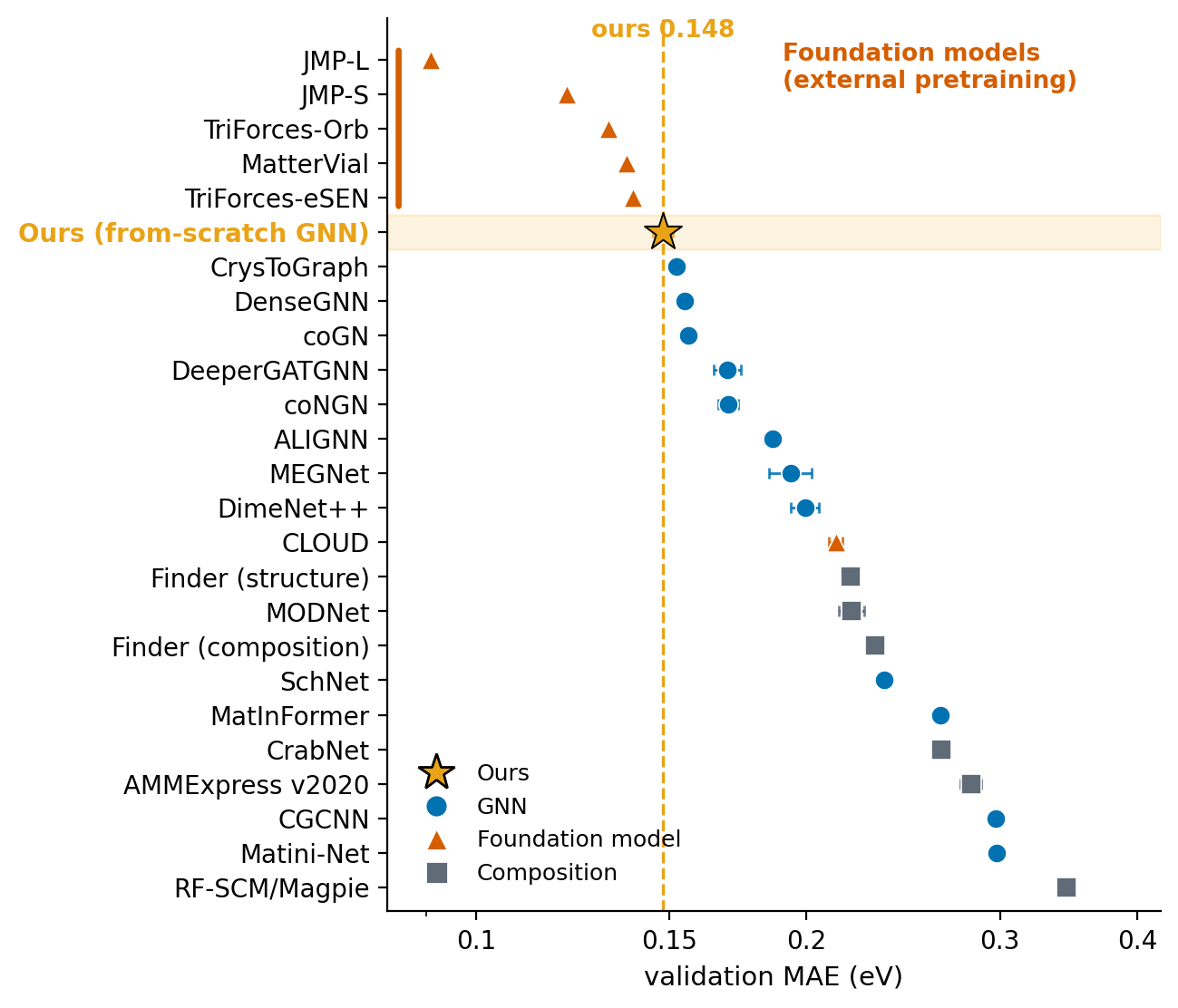}
\caption{Mean absolute error (eV, log scale, lower is better) on the \texttt{mp\_gap} task for the
agent's best model (gold star), compared against published methods at their reported scores from the
public MatBench leaderboard or source paper \citep{dunn2020matbench}. Blue circles mark from-scratch
single-task neural networks, most of them crystal graph networks: coGN and coNGN \citep{ruff2023cogn}, CrysToGraph
\citep{crystograph2024}, DenseGNN \citep{densegnn2025}, DeeperGATGNN \citep{omee2022deepergatgnn},
ALIGNN \citep{choudhary2021alignn}, MEGNet \citep{chen2019megnet}, DimeNet++
\citep{gasteiger2020dimenetpp}, SchNet \citep{schutt2018schnet}, MatInFormer \citep{huang2023matinformer},
CGCNN \citep{xie2018cgcnn}, and Matini-Net \citep{lee2024matini}. Orange triangles mark foundation-model
methods that import external large-scale pretraining: JMP \citep{shoghi2023jmp}, TriForces
\citep{triforces2026}, MatterVial \citep{mattervial2026}, and CLOUD \citep{xu2026cloud}. Grey squares
mark composition or other baseline models: Finder \citep{ihalage2022finder}, MODNet
\citep{debreuck2021modnet}, CrabNet \citep{wang2021crabnet}, AMMExpress v2020 \citep{dunn2020matbench},
and RF-SCM/Magpie \citep{ward2016magpie}.}
\label{fig:standing}
\end{figure}

% NOTE: what we did and the headline result -- the test field and its scale, the minimal loop, the surprising result, and the reading that the gain came from recombining existing methods of two kinds.
We therefore use the MatBench benchmark \citep{dunn2020matbench} as a test field for this question,
focusing on its band-gap task: more than 100{,}000 crystals and more than twenty published models with
reported scores. We minimize the agent's workflow: each experiment edits a training script, costs real
compute, and returns only a single held-out number, with no human in the loop
\citep{karpathy2026autoresearch}. The result is surprising: with no materials-specific design, the agent
built the most accurate model trained without external pretraining, ahead of all seventeen such
from-scratch models reported for the task (Figure~\ref{fig:standing}). Analyzing the final model shows that
this improvement came from recombining existing methods rather than inventing new ones. The methods it
recombined are of two kinds, those already existing in standard crystal neural networks and those borrowed from
other areas of machine learning, and the result follows from combining them rather than from any single
one (Figure~\ref{fig:rediscovery}, Table~\ref{tab:novelty}).

% NOTE: the second contribution -- the loop's failure mode, the same search problem in other agents and their approaches, our simpler crowding score, and the measured result on a second task.
We also find a limitation of these loops: the agent keeps making small changes to one model and stops
improving, instead of exploring a different approach. Other autonomous research agents address the same
problem with different strategies: for example, an evolutionary population of programs in AlphaEvolve
\citep{novikov2025alphaevolve}, Monte Carlo tree search over candidate solutions in AB-MCTS
\citep{inoue2025abmcts}, and agentic tree search in the AI Scientist \citep{yamada2025aiscientistv2}. Here we use a simpler method:
a crowding score that measures how much the best model so far makes the same mistakes as other equally
accurate models, and asks the agent to try a different approach when that score is high. With the crowding
score, on a second task (shear-modulus prediction), the loop reaches a mean absolute error of $0.0699$ in
$\log_{10}$ GPa by experiment $40$, while the same loop without it never gets below $0.0721$ across all
$160$ experiments (Section~\ref{sec:crowded}).

\section{Experimental Setup}
% NOTE: the main task mp_gap, then the published crystal-graph-network scores we compare against (coGN and the two from-scratch networks that report lower error).
The main task we used is \texttt{mp\_gap} \citep{dunn2020matbench}: predict the electronic band gap in eV
from a crystal structure, scored as the mean absolute error over a fixed five-fold partition of more than
$100{,}000$ crystals. The standard reference on this task is coGN \citep{ruff2023cogn}, a crystal graph
network built on an optimized way of connecting the atoms into a graph, at
$0.1559$ eV. Two newer crystal graph networks report lower error:
CrysToGraph at $0.1522$ eV \citep{crystograph2024}, and DenseGNN, which adds dense connections between
layers, at $0.1548$ eV for its Voronoi variant \citep{densegnn2025}. These are among the published models we compare against
(Figure~\ref{fig:standing}).

% NOTE: how the loop works -- the Karpathy edit/train/keep-or-drop cycle, our tau filter, the per-run budget and hardware, the finalization run, and the fixed scripts that make every candidate comparable.
Our loop follows the autonomous-research setup of \citet{karpathy2026autoresearch}: the agent edits the
training code, the edited model is trained under a fixed budget, the change is kept or dropped by its
measured accuracy, and the loop repeats. In addition to this conventional procedure, we require a change
to lower the held-out mean absolute error by more than $\tau = 0.002$ eV to count as a real improvement,
not run-to-run noise.
Each training run is capped at $1{,}500$ seconds (25 minutes) on a single RTX~2070~SUPER GPU, and each
candidate model is limited to about five million parameters, so every experiment finishes quickly and the
loop can run hundreds of them.
% source: Claude Code / Claude Opus 4.8 has no formal paper to cite; add a footnote or URL here if a reference is wanted.
We use Claude Code as the agent, running Claude Opus 4.8 at its highest thinking-effort setting (\texttt{max}).
During the search, each candidate is ranked on a single held-out fold (fold~0).
After the loop ends, we take the lowest-error model as the final model and
train it on the official five-fold split, $9{,}000$ seconds per fold, to report its accuracy fully trained
rather than under the search budget. Throughout the loop, the agent edits only the model code, while data
loading, the held-out fold, and the error computation are run by fixed scripts the agent cannot change, so
every candidate is judged the same way.

% NOTE: the second task gvrh used in Section 5 -- what it predicts, why its small dataset lets us cut the budget and iterate on the workflow, and that the rest of the loop matches the band-gap task.
In Section~\ref{sec:crowded} we use a second task, \texttt{log\_gvrh}, which predicts the
base-ten logarithm of the shear modulus in GPa. This task has far fewer crystals ($\sim$11{,}000), so
each model trains quickly and we cut the per-run budget to $180$ seconds. The smaller budget lets us
run many experiments cheaply and focus on developing the agent's workflow.
Apart from the task, its data, and this budget, the loop and the fixed scripts are the same as for
the band-gap task.

\section{A competitive model assembled from known methods}
% NOTE: open the section with the competence the rest of the paper qualifies: the headline five-fold result read off Figure 1, scoped honestly (most accurate without external pretraining, not a state-of-the-art claim).
Figure~\ref{fig:standing} compares the loop's final model with the published models for the task, each
shown at its reported mean absolute error on the official benchmark. The model reaches a mean absolute
error of $0.1480 \pm 0.003$ eV, ranking 6th in the figure. The five models ahead of it (orange triangles)
all rely on external pretraining. Four are foundation models pretrained on additional data, JMP-L and
JMP-S \citep{shoghi2023jmp} and TriForces-Orb and TriForces-eSEN \citep{triforces2026}. The fifth,
MatterVial \citep{mattervial2026}, imports features from other pretrained models. Among the models trained
from scratch on this task alone, the loop's model (marked as the gold star) is more accurate than every
expert-designed one.

\begin{figure}[t]
\centering
\includegraphics[width=0.82\linewidth]{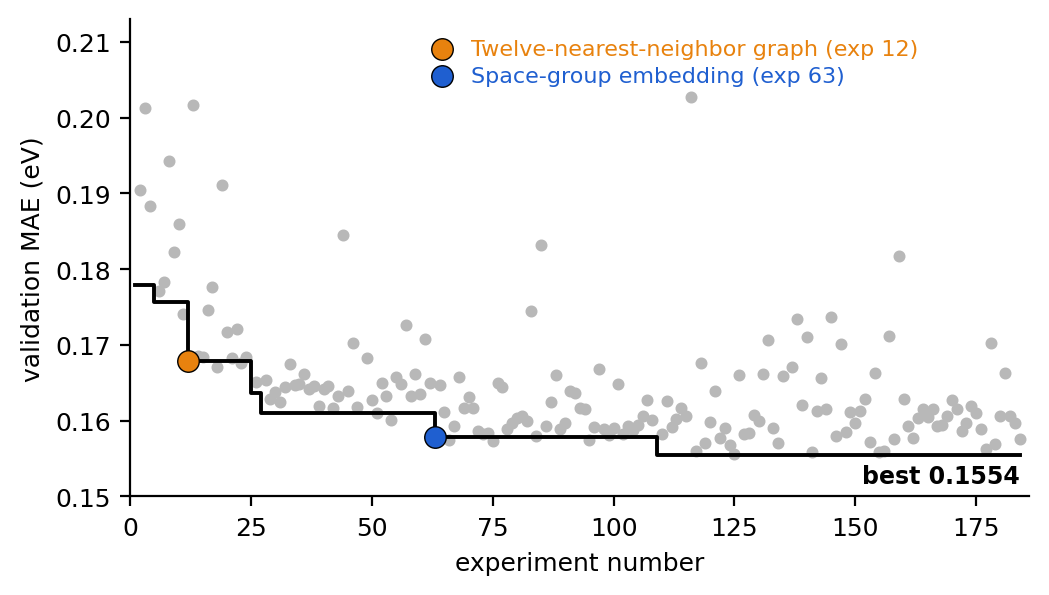}
\caption{The lowest validation MAE reached (eV, solid black line) against experiment number on a single
fold of the \texttt{mp\_gap} task. Each dot is one experiment at its MAE, and the grey dots are the
disregarded experiments. Two accepted changes are marked as examples: a twelve-nearest-neighbor graph
(orange, experiment 12) and a space-group embedding (blue, experiment 63). The final best model reaches
$0.1554$, and falls to $0.148$ with a longer training budget and five-fold validation.}
\label{fig:rediscovery}
\end{figure}

% NOTE: run-in claim header; lead with the traced descent and its measured span, scope the claim (reference implementations were available, so it is about which edits the accept rule kept and in what order, not invention).
Figure~\ref{fig:rediscovery} shows the agent's exploration over 184 experiments. The solid line shows the
best held-out MAE kept so far. The line decreases when an experiment's MAE is at least $0.002$ eV below the
current best, for example the bond-angle and connectivity change at experiment~12 (to $0.1679$ eV) and the
space-group symmetry change at experiment~63 (to $0.1578$ eV). The grey dots are the
attempts that did not improve on the best and were discarded. Seven kept changes in all bring the model
from its untuned baseline near $0.18$ eV to its best model at $0.1554$ eV, listed in order in
Table~\ref{tab:keeps}. Two are increases in the model width with no associated method. The rest are
representational changes, each matching a method already in the crystal-network literature. The largest
single gain adds bond-angle messages through a line graph (a graph whose nodes are the edges of the crystal
graph, so the model can use the angles between bonds), the construction used by ALIGNN
\citep{choudhary2021alignn}; DimeNet++ \citep{gasteiger2020dimenetpp} uses the same bond-angle
information through directional message passing. The same change also adds an optimized
nearest-neighbour connectivity in the style of coGN \citep{ruff2023cogn}. The later changes add element-pair features on each edge, a
space-group embedding for crystal symmetry in the manner of MEGNet's global features
\citep{chen2019megnet}, and a wider Gaussian basis for the interatomic distances. Together, these changes
make a single crystal graph network that recombines methods from several published models, with its full
architecture in Appendix~\ref{app:champion}.

\begin{table}[t]
\centering
\caption{Novelty classification of every distinct method the agent proposed, split by whether it was kept or
discarded. Each method is one of: \emph{rediscovery} (it appears in a crystal-GNN paper),
\emph{cross-domain} (absent from the crystal-GNN set but used in other ML papers), or \emph{novel} (in
neither). Each method is a real algorithm change, and pure hyperparameter sweeps are excluded.}
\label{tab:novelty}
\begin{tabular}{lcccc}
\toprule
Methods proposed & Rediscovery & Cross-domain & Novel & Total \\
\midrule
Accepted  & 5  & 0 & 0 & 5  \\
Discarded & 20 & 5 & 0 & 25 \\
\midrule
Total     & 25 & 5 & 0 & 30 \\
\bottomrule
\end{tabular}
\end{table}

% NOTE: run-in claim header; lead the body with the result the table yields (invented nothing), pointing to the table for the categorization method rather than re-describing it.
To track where the agent's ideas came from, we analyzed every attempt it made, including the discarded
ones, and classified them into three classes by the novelty of each idea. First we identified the real
algorithm changes, setting aside pure hyperparameter sweeps such as width or learning-rate changes. This
left 30 distinct methods. We then compared each against a reference set of 21 papers from this field, the
published models of Figure~\ref{fig:standing} among them (Table~\ref{tab:novelty}, with per-change sources
in Appendix~\ref{app:provenance}). A method is a rediscovery if it already appears in one of the crystal
graph networks, a cross-domain import if it is absent from them but taken from feature-based materials
machine learning or general deep learning, and novel if it appears in neither. All 5 accepted changes are
rediscoveries, and 20 of the 25 the loop tried and reverted are also rediscoveries. The remaining 5 are cross-domain imports. They add an inverse-distance Coulomb edge, an
element-product edge, and a per-atom neighbor-composition descriptor, train with a robust regression loss,
and apply stochastic depth. None of the 30 is novel. The competitive model of Figure~\ref{fig:standing} is
therefore built by recombining known methods alone.

\section{The model does not transfer to other tasks}
% NOTE: para A (no bold, no figure pointer) poses the question and names + justifies the four tasks, per cz sentence logic.
The band gap is only one of many properties that can be predicted from a crystal's structure. One may therefore ask whether our model, optimized for band-gap prediction, also predicts other properties well. To answer this, we keep the final model's architecture and hyperparameters fixed and retrain it on four other MatBench tasks: elastic-modulus prediction (\texttt{log\_gvrh} and \texttt{log\_kvrh}) and formation-energy prediction (\texttt{mp\_e\_form} and \texttt{perovskites}). We chose these four because each predicts a property from a crystal's structure, the same kind of task as the band gap.

\begin{figure}[t]
\centering
\includegraphics[width=0.82\linewidth]{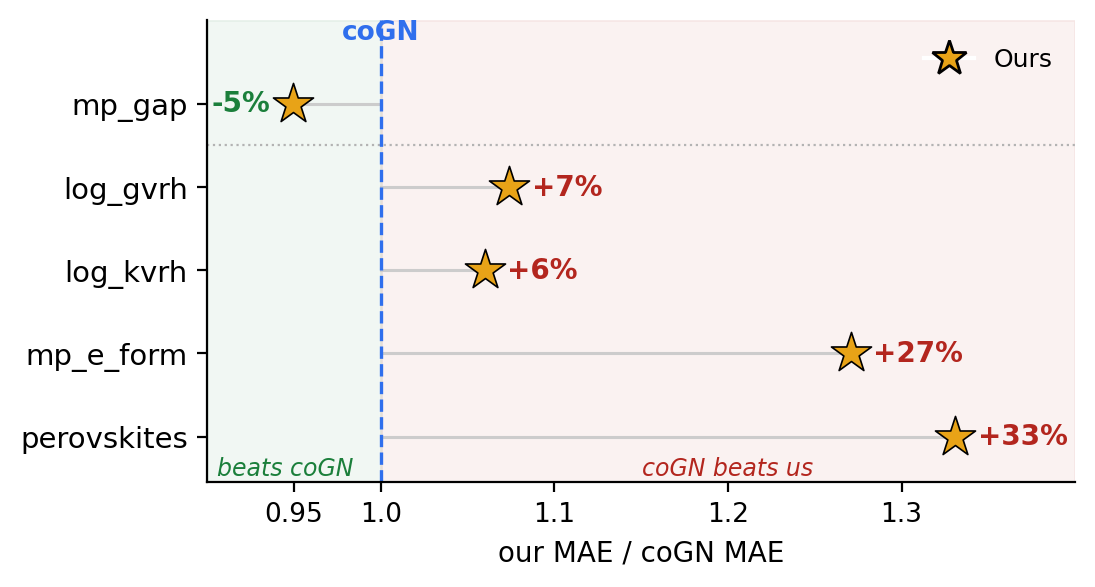}
\caption{The best model's mean absolute error (gold stars) relative to coGN's, on \texttt{mp\_gap}
(top) and the four other tasks it was retrained on: \texttt{log\_gvrh}, \texttt{log\_kvrh},
\texttt{mp\_e\_form}, and \texttt{perovskites}. The horizontal axis is the best model's MAE divided by
coGN's MAE on the same task, so the dashed line at $1.0$ is coGN: a star left of it (green) means the
best model beats coGN, and right of it (red) means coGN beats it, with the percentage difference
labeled by each star.}
\label{fig:transfer}
\end{figure}

% NOTE: para B -- Figure 3 frames the coGN comparison; why coGN (a strong, single general design optimized on log_gvrh, applied unchanged); coGN's transferability and its reason from the coGN paper; the measured result; our per-task leaderboard rank; the conclusion.
Figure~\ref{fig:transfer} compares our model's mean absolute error with that of coGN, an expert-designed crystal graph network, on the band gap and the four other tasks. We compare against coGN because it is one of the strongest models across these tasks and is a single general design: its authors found one architecture and one set of hyperparameters by optimizing on a single task, \texttt{log\_gvrh}, then applied them unchanged to every other task with no per-task tuning \citep{ruff2023cogn}. They report that this one configuration carries over to the other MatBench datasets and tasks, and attribute its accuracy to an optimized choice of which atoms are joined in the graph and to deeper message-passing layers. As shown in Figure~\ref{fig:transfer}, our model beats coGN on the band gap but loses to it on all four other tasks, by six to thirty-three percent. On those four tasks the retrained model places 5th, 4th, 4th, and 8th respectively among the established MatBench submissions, an ordinary mid-leaderboard standing rather than a leading one like coGN. These results indicate that the loop produced a model tuned to the band gap: its standing comes from automated search and tuning on a single task, not from a design that holds across tasks. An expert-designed model like coGN instead rests on a general design choice grounded in crystal structure, how atoms are connected in the graph, which is why its single configuration transfers where our search-tuned model does not.

\section{Getting the agent to try new models}
\label{sec:crowded}
\begin{figure}[t]
\centering
\includegraphics[width=0.82\linewidth]{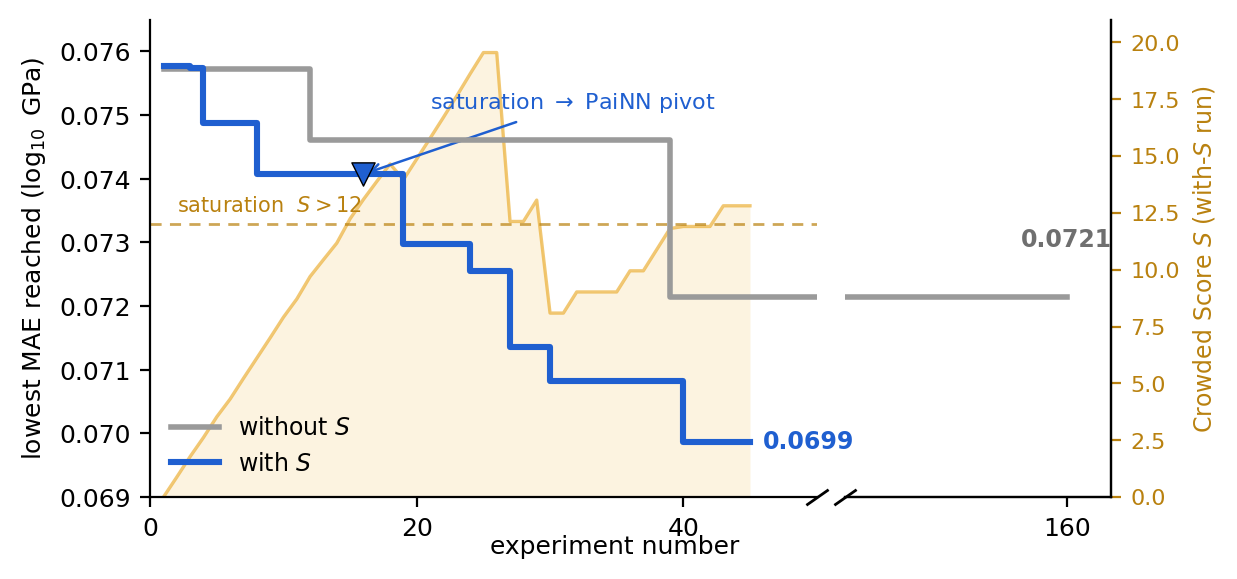}
\caption{The lowest held-out MAE reached, against experiment number, for two runs on the
\texttt{log\_gvrh} task, one without the crowding score $S$ (grey line) and one with it (blue line),
where the agent is prompted to try a different model when $S$ is high. The blue triangle marks where the
agent switched from a CGCNN-style convolutional network to a PaiNN-style equivariant network. The right axis
shows $S$, which measures how much the best model makes the same mistakes as other equally accurate
models (defined in the main text).}
\label{fig:crowded}
\end{figure}

% NOTE: the failure mode and the open question -- the loop refines one model and stops improving (mp_gap, then more clearly on log_gvrh: 0.0721 floor, ~120 stuck experiments that are small within-family changes, bigger changes rejected on a worse first try), then the refine-vs-explore balance as a known open problem with two cited approaches.
In the \texttt{mp\_gap} task, the AI agent lowered its error quickly at first and then stopped: its best model reached its lowest error after about one hundred experiments and did not improve over the rest of the run. The agent kept refining one model it had already tuned well, rather than trying a different kind of model. This pattern is clearer on a second task, \texttt{log\_gvrh}, which predicts the base-ten logarithm of the shear modulus in GPa. Seeded with the best \texttt{mp\_gap} model retrained on this task at about $0.076$, the agent lowered the error to $0.0721$ by about experiment $40$, then ran more than one hundred further experiments without lowering it again (Figure~\ref{fig:crowded}, grey). It stayed there because the agent kept making small changes to the same model, adjusting its learning rate, width, depth, or optimizer, and never tried a different one, such as a different graph structure or feature embedding. How to explore the space of possible solutions efficiently is a central question for automated research agents, and different systems answer it differently. AlphaEvolve \citep{novikov2025alphaevolve} uses an evolutionary database based on MAP-Elites and island models, and AB-MCTS \citep{inoue2025abmcts} uses Thompson sampling to choose between a new candidate and a refinement.

% NOTE: our simpler method vs the cited approaches: define the crowding score (equation kept), give its physical meaning and the S>12 switch rule, state it is harness-computed and the only thing added; then the controlled A/B off Figure 4 (cross the threshold at 15, pivot to PaiNN at 16, worse first try, beat the convolution best at 19, reach 0.0699 at 40). Merges the old controlled-comparison and figure-mechanism paragraphs.
Rather than rebuild the search like those systems, here we apply a simpler approach. We compute a single quantity, the crowding score $S(e)$ for each committed model $e$, to measure how much model $e$ repeats the models already tried and to prompt the agent to propose a new idea when $S$ is high. The score is defined as
\begin{equation}
S(e) \;=\; \sum_{\substack{f \neq e \\ |\mathrm{MAE}_e - \mathrm{MAE}_f| \,<\, \tau}} \max\!\bigl(0,\; \rho(\mathbf{r}_e, \mathbf{r}_f)\bigr),
\qquad \tau = 0.004,
\label{eq:crowded}
\end{equation}
where $\mathbf{r}_e = \hat{\mathbf{y}}_e - \mathbf{y}$ is model $e$'s vector of per-crystal errors on a fixed held-out set, $\mathrm{MAE}_e = \mathrm{mean}(|\mathbf{r}_e|)$ is its mean absolute error, $\rho$ is the Pearson correlation between two such vectors, and the sum runs over every other committed model whose error is within $\tau$ of $e$'s. A model's residual vector is its fingerprint, so two highly similar models produce residual vectors that are highly correlated, giving a high $\rho$. When the loop only makes small changes to the best model, it produces many new models that are highly similar to it. Each of these has a high $\rho$ with the best model, so the best model's score $S$ rises quickly. Once the best model has $S > 12$, we tell the agent to stop refining it and build a different kind of model. The score is computed by the harness from residuals the agent never sees, and it is the only thing added to the loop. With the score fed back, the loop's convolution models grew crowded enough to pass the threshold at experiment $15$, and the agent switched to a PaiNN-style equivariant network, one that carries directional vector features alongside the usual scalar ones \citep{schutt2021painn}, at experiment $16$. That first version scored worse than the convolution models, but after a few refinements it passed their best at experiment $19$ and reached $0.0699$ by experiment $40$, below the $0.0721$ the run without the score never passed despite running more than three times as many experiments (Figure~\ref{fig:crowded}).

\section{Related work}
% NOTE: thin positioning paragraph - the published ML models for this task are the comparison set. Point to Figure 1, defer references to its caption, and close on the scoped standing (lowest reported error among models without external pretraining).
\textbf{Band-gap models for this task.} Predicting the band gap from crystal structure is an
established machine-learning task with many published models (Figure~\ref{fig:standing}), ranging
from message-passing crystal graph networks to foundation models pretrained on additional data.
Among the models trained without external pretraining, the loop's model reaches the lowest reported
error. References for the compared models are listed in that figure's caption.

% NOTE: position the loop in the autonomous-research-loop line, then turn on two honest contrasts: (1) prior loops were run on ML/coding/math (general code changes), ours on a materials task that calls for domain knowledge; (2) materials LLM agents do discovery/synthesis, not model optimization, so this was uncharacterized.
\textbf{Autonomous research loops.} LLM agents that propose, run, and keep code changes against a
held-out metric have been built for machine-learning engineering: \citet{jiang2025aide} frame the work as
a tree search over training code, and \citet{mlebench2025agents} study the search policy and operator
set that such loops use. Related systems automate a wider research cycle \citep{lu2024aiscientist} or
evolve programs that improve on the best known algorithm
\citep{romeraparedes2024funsearch,novikov2025alphaevolve}. These systems were demonstrated on
machine-learning, coding, and mathematics benchmarks, where a helpful change is a general code or
training adjustment. We instead apply such a loop to a materials-property benchmark, which calls for
domain knowledge of crystal structure, chemical bonding, and space-group symmetry rather than a
general code change. Autonomous agents in materials science have largely proposed new compounds and
synthesis routes \citep{jia2024llmatdesign,szymanski2023alab} rather than optimizing a
property-prediction model, so what such a loop converges on for this task has not, to our knowledge,
been characterized.

\section{Conclusion}
% NOTE: Purpose -- one paragraph (cz's structure): pose the question, give the result, restate it as a title callback, name the two limitations, give the crowding score as the fix for the second, then a forward note. Moves: (1) question -- can a coding-agent loop produce a model competitive with expert-designed ones; (2) result -- on the band-gap task the loop built the most accurate model trained from scratch, ahead of every expert-designed one, by assembling known methods rather than inventing one; (3) callback to the title -- a simple autonomous LLM research loop can optimize an expert-designed model and improve its accuracy; (4) two limitations -- (a) the optimized model does not transfer to other tasks, unlike physically-reasoned coGN, (b) the agent stops at small adjustments instead of a different approach; (5) the crowding score as the fix for (b), a scalar for repeated errors among equally accurate models, prompting a different class of model when high, after which the loop reached a lower error than the run without it; (6) forward -- applying the same loop to other graph-network families, such as machine-learning interatomic potentials and foundation models for materials, is future work. Numberless; formal register; uses 'best model', not 'champion'. Polished via the rule-review loop; cz sharpened the opening, dropped the former 'room to improve' implication, and added the title callback.
In this work, we investigate whether a general-purpose coding agent, run as an autonomous loop on a
mature materials benchmark, can produce a model competitive with expert-designed ones. On the band-gap
task, the loop built the most accurate model trained from scratch, ahead of every expert-designed one, by
assembling existing methods rather than inventing a new one. This result indicates that a simple
autonomous LLM research loop can optimize an expert-designed model and improve its accuracy. Two
limitations remain: (a) the model the agent optimized does not transfer to other tasks, unlike an
expert-designed model built from physical reasoning such as coGN, and (b) the agent tends to stop
improving, making small adjustments rather than exploring a different approach. For the second limitation,
we propose the crowding score, a scalar that measures the extent to which the current best model repeats
the errors of other equally accurate models. When this score is high, we prompt the agent to develop a
different class of model. With the crowding score, the loop reached a lower error than the same loop run
without it. Applying the same loop beyond band-gap prediction, to other graph-network model families such
as machine-learning interatomic potentials and foundation models for materials, is a direction for future
work.

\subsubsection*{Acknowledgments}
This work was supported by the Army Research Office (EMSS, W911NF-25-1-0225).

\bibliographystyle{colm2026_conference}
\bibliography{refs}

\appendix

\section{Final model architecture}
\label{app:champion}
% NOTE: name the model and tie each component to the representational primitive and field method of Finding 1.
The final model (about $503{,}617$ parameters) is a crystal graph network with width $120$ and four
message-passing layers. A crystal is converted to a graph with a $5.0$\,\AA{} cutoff and a
$k$-nearest-neighbour cap of $12$ (the connectivity primitive). Each edge carries a $32$-component
Gaussian radial basis of the interatomic distance together with the element pair of its endpoints (the
distance and composition primitives). One ALIGNN-lite update passes angle-aware messages between bonds
that share an atom, using a line graph and an $8$-component basis of the bond-angle cosine (the angle
primitive). Four CGCNN-style gated convolutions \citep{xie2018cgcnn} update the atom states. A global term
concatenates six lattice and density statistics with a $16$-dimensional embedding of the space group, one
of $231$ (the symmetry primitive). Readout pools the atom states by mean and max, appends the global term,
and regresses the band gap. Weights are trained with Adam under a cosine schedule, and predictions use an
exponential moving average of the weights.

\section{Kept changes and their provenance}
\label{app:provenance}
% NOTE: the kept-change chain (MAE, the primitive each changes, and the prior crystal-network method each rediscovers) merged with the search cost; then the cross-domain methods tried and reverted. The redundant accept-threshold and per-run-budget text is dropped (it is in the setup).
Table~\ref{tab:keeps} lists the kept changes of the reported search, in order. Table~\ref{tab:crossdomain}
lists the cross-domain methods the agent tried and reverted.

\begin{table}[h]
\centering
\small
\begin{tabular}{c c >{\raggedright\arraybackslash}p{3.5cm} >{\raggedright\arraybackslash}p{1.9cm} >{\raggedright\arraybackslash}p{3.5cm}}
\toprule
Exp. & MAE (eV) & Change & Primitive & Prior source (rediscovery) \\
\midrule
1   & 0.1780 & seed baseline & distances & -- \\
5   & 0.1757 & width $64 \to 96$ & capacity & -- \\
12  & 0.1679 & $k$-NN cap $K{=}12$ and ALIGNN-style angle update & connectivity, angles & CGCNN \citep{xie2018cgcnn}; ALIGNN \citep{choudhary2021alignn}, DimeNet++ \citep{gasteiger2020dimenetpp} \\
25  & 0.1636 & element-pair features in the edge MLP & composition & MEGNet \citep{chen2019megnet} \\
27  & 0.1610 & width $96 \to 112$ & capacity & -- \\
63  & 0.1578 & space-group embedding and width $120$ & symmetry & CLOUD \citep{xu2026cloud}, Automatminer \citep{dunn2020matbench} \\
109 & 0.1554 & wider Gaussian radial basis (final model) & distances & SchNet \citep{schutt2018schnet} \\
\bottomrule
\end{tabular}
\caption{Each kept change of the reported search, in order, with the experiment that introduced it, its
MAE, the representational primitive it changed, and the prior crystal-network method it rediscovers. A
dash marks the seed baseline and the width increases, which are not methods.}
\label{tab:keeps}
\end{table}

\begin{table}[h]
\centering
\small
\begin{tabular}{c >{\raggedright\arraybackslash}p{4.6cm} >{\raggedright\arraybackslash}p{7.0cm}}
\toprule
Exp. & Cross-domain import (tried, reverted) & Prior source \\
\midrule
42 & Per-atom neighbor-composition descriptor & MatterVial \citep{mattervial2026} \\
92 & Element-product edge & MatterVial \citep{mattervial2026}, CrabNet \citep{wang2021crabnet} \\
126 & Robust regression loss & CrabNet \citep{wang2021crabnet}, FINDER \citep{ihalage2022finder} \\
151 & Inverse-distance Coulomb edge & MatterVial \citep{mattervial2026} \\
169 & Stochastic depth & general deep learning \citep{huang2016stochasticdepth} \\
\bottomrule
\end{tabular}
\caption{The five cross-domain methods the agent tried and reverted, with the experiment that introduced
each and its prior source.}
\label{tab:crossdomain}
\end{table}

\section{Prompts with crowding score}
\label{app:crowding}
% NOTE: show program.md, the full agent prompt for the crowding-score run of Section 5. One sentence saying what it is; the crowding-score rule and equation are already in the main text, so they are not repeated here. The program is boxed verbatim.
Below is \texttt{program.md}, the full prompt that defines the agent's loop in the crowding-score run of
Section~\ref{sec:crowded}.

\begin{framed}
\small
\noindent\textbf{autoresearch -- log\_gvrh (novelty-driven, representation-first)}\par\smallskip
You are a creative materials-science ML researcher. Goal: the LOWEST \texttt{mae:} (log10 GPa) on matbench\_log\_gvrh, the official held-out fold, under a fixed compute budget. You edit ONE file (\texttt{train.py}), run it, and commit it. There is NO reset -- every model is committed and KEPT IN HISTORY so novelty.py can fingerprint it. But you always REFINE FROM THE BEST model so far (lowest mae), never from whatever was committed last (see "Each round"). The best model is recovered at the end. Loop until interrupted.\par\smallskip
\smallskip\noindent\textbf{Setup (once)}\nopagebreak\par\smallskip
\begin{enumerate}[leftmargin=1.4em,itemsep=2pt,topsep=2pt,parsep=0pt]
\item Read \texttt{prepare.py} (READ-ONLY: data, official fold, private TEST labels, the MAE metric -- never edit it, never read \texttt{data/*} directly) and \texttt{train.py} (the ONLY file you edit). \texttt{ref/} has real installed reference code if you want it (matgl line-graph / three-body utils, coGN / coNGN / ALIGNN) -- take ideas, do not copy; you are not limited to known approaches.
\item Create \texttt{results.tsv} with the header if missing (see Logging).
\end{enumerate}
\smallskip\noindent\textbf{You are a scientist, not a tuner}\nopagebreak\par\smallskip
Your job is to understand why the current model fails and to invent a better one -- any representation, architecture, or inductive bias you think could work. Do not limit yourself to known or conventional approaches; pursue whatever ideas you find genuinely promising, including ambitious or unconventional ones. Depth of understanding over number of edits; aim for novelty.\par\smallskip
\smallskip\noindent\textbf{Each round}\nopagebreak\par\smallskip
\begin{enumerate}[leftmargin=1.4em,itemsep=2pt,topsep=2pt,parsep=0pt]
\item STUDY: read \texttt{novelty\_report.md} -- \texttt{S(a)} (how behaviorally crowded the model you develop from -- the best method so far -- is: how many same-scoring models already make the same predictions as it) and whether it is SATURATED (\texttt{S(a) \textgreater{} 12}). The "most crowded models" list shows redundancy across the whole family. When you need to diverge, read the \texttt{results.tsv} descriptions to see which mechanisms already exist.
\item INTERPRET: think through what the best model is failing to capture, and why. Reason from the physics of elasticity and bonding where it helps. Understanding first; edits second.
\item DECIDE:
\begin{itemize}[leftmargin=1.2em,itemsep=1pt,topsep=1pt,parsep=0pt]
\item NOT saturated and you have a concrete improvement $\rightarrow$ refine the best model.
\item SATURATED (\texttt{S(a) \textgreater{} 12} -- the family keeps building near-duplicates that score the same and make the same mistakes) $\rightarrow$ STOP refining; this mechanism is exhausted. Invent a fundamentally different model, one NOT in the crowded clusters \texttt{novelty\_report.md} names. A coupled change touching several parts of \texttt{train.py} at once is expected.
\end{itemize}
\item IMPLEMENT: FIRST run \texttt{bash checkout\_best.sh} -- it restores \texttt{train.py} to the best (lowest-mae) model so far, so you build on the STRONGEST model, not whatever was committed last. It changes only the working file; every commit stays in history. Run it BEFORE editing -- running it after would discard your edit. THEN edit ONLY \texttt{train.py} on top of the restored best.
\item RUN, then COMMIT, then LOG -- the order matters (see below).
\end{enumerate}
\smallskip\noindent\textbf{Run / commit / log (order matters)}\nopagebreak\par\smallskip
\begin{lstlisting}
bash checkout_best.sh                                  # FIRST: restore train.py to the best model
# ... now edit train.py ...
source .venv/bin/activate
python train.py > run.log 2>&1
grep "^mae:\|^num_params:\|^peak_vram_mb:" run.log     # empty mae => crash (tail run.log)
git add -A && git commit -m "<one-line reasoning: mechanism + hypothesis>"
\end{lstlisting}
\begin{itemize}[leftmargin=1.2em,itemsep=2pt,topsep=2pt,parsep=0pt]
\item RUN, then COMMIT: the run stages this model's fingerprint. THEN commit -- the post-commit hook keys the fingerprint by the new commit and rewrites \texttt{novelty\_report.md}. Never skip the commit.
\item After committing, append a \texttt{results.tsv} row: \texttt{git rev-parse --short HEAD}, the mae, peak\_vram, a one-line mechanism + hypothesis description. That description is your durable memory -- make it specific (your fuller reasoning goes in the commit message).
\end{itemize}
\smallskip\noindent\textbf{CAN / CANNOT}\nopagebreak\par\smallskip
CAN (all in \texttt{train.py}): featurization (cutoff, connectivity, line graphs / bond angles, node / edge / global features, multiple graphs per crystal), model, optimizer, schedule, training loop, batch size, readout. CANNOT: edit \texttt{prepare.py}; change the data / fold split / TEST labels / \texttt{prepare.evaluate\_mae} metric; read \texttt{data/*} or \texttt{fingerprints/*} directly; install packages (import what is already installed); exceed 5,000,000 params (print \texttt{num\_params}); touch \texttt{run.sh}, the git hooks, \texttt{novelty.py}, \texttt{checkout\_best.sh}, tmux, or any process but \texttt{python train.py}; end this session.\par\smallskip
\smallskip\noindent\textbf{Hard constraints}\nopagebreak\par\smallskip
\begin{enumerate}[leftmargin=1.4em,itemsep=2pt,topsep=2pt,parsep=0pt]
\item ENVELOPE -- matched, not scaled: $\leq$ 5M params, single GPU, FP16 ok. A win must be algorithmic / representational. Heavier graphs (line / three-body) train fewer steps in the fixed budget -- that is the rule; if a richer model is close but under-trained, shrink it rather than drop the idea.
\item FIXED BUDGET -- each run trains \texttt{prepare.TIME\_BUDGET} s (180), then evaluates.
\item FIXED SEED -- \texttt{train.py} seeds RNG at the top so one run is a fair comparison. Keep it.
\item FEAT CACHE -- \texttt{featurize()} caches per \texttt{FEAT\_KEY} under \texttt{prepare.cache\_dir(FEAT\_KEY)}. CHANGE \texttt{FEAT\_KEY} whenever featurize() output changes, or you read a stale cache. \texttt{feat\_seconds} is printed and does NOT count against the budget.
\end{enumerate}
\smallskip\noindent\textbf{Logging (results.tsv, TAB-separated)}\nopagebreak\par\smallskip
\begin{lstlisting}
commit	mae	memory_gb	status	description
\end{lstlisting}
short-hash \textbar{} mae (0.000000 if crash) \textbar{} peak\_vram\_mb/1024 \textbar{} keep (always -- no reset) or crash \textbar{} one-line mechanism + hypothesis.\par\smallskip
\smallskip\noindent\textbf{Baseline}\nopagebreak\par\smallskip
Iteration 1 = the seed \texttt{train.py} AS IS: run once, commit, log (status keep). Change nothing before it is recorded. If \texttt{results.tsv} already lists kept models, the baseline is done -- do NOT repeat it; continue the loop. \texttt{novelty\_report.md} is empty until after the first commit -- that is expected.\par\smallskip
\smallskip\noindent\textbf{Crashes}\nopagebreak\par\smallskip
Trivial (typo / import) $\rightarrow$ fix and re-run. Fundamentally broken $\rightarrow$ log mae \texttt{0.000000}, commit anyway so history stays complete, move on. NaN pooled stats are common -- clamp variances, avoid dividing by zero atom counts. Line-graph / three-body code is crash-prone; build incrementally, keep \texttt{FEAT\_KEY} in sync.\par\smallskip

\end{framed}

\end{document}